\documentclass[12pt]{article}
\pdfoutput=1
\usepackage{amssymb,amsmath}
\usepackage{amsbsy}
\usepackage{amsfonts}
\usepackage[dvips]{graphicx}
\usepackage{cancel}
\usepackage{enumerate}
\usepackage[colorlinks=true]{hyperref}
\newcommand{\AddrSanto}{
  {\it Universidad Santo Tomas, Bogot\'a, Colombia}}
\newcommand{\AddrUNAL}{
  {\it Departamento de F\'{\i}sica, Universidad Nacional de Colombia, Bogot\'a, Colombia}}

\title{Charged Lepton mixing processes in 331 Models}

\author{J. M. Cabarcas$^1$, J. Duarte$^2$, J.-Alexis Rodriguez$^2$\\
1.\AddrSanto \\ 2.\AddrUNAL}
\date{}
\begin{document}
\maketitle{}

\begin{abstract}

Processes $\tau\rightarrow l\gamma$,  $\tau\rightarrow lll$ with $l= e,\mu$ and $\mu(\tau) \to e (\mu) \gamma$ are evaluated
in the framework of a 
model based on the extended symmetry gauge $SU(3)_c \otimes SU(3)_L \otimes U(1)_Y$
 with a leptonic sector consistent of five triplets. Lepton flavor violating processes are allowed at tree level in this model through the new $Z'$ gauge boson.
 We obtained bounds for the mixing angles in the leptonic sector of the model, considering the experimental measurements
 of the processes from the BELLE and the BABAR collaborations.

\end{abstract}

\section{Introduction}

In the framework of the standard model (SM) of high energy physics there are many unclear issues 
that  require extensions of the theory in the local symmetry and in the particle spectrum.
One of these issues is the flavor puzzle: why there are three families of fermions and why they have their peculiar structure of masses and mixings.
There are many studies of these problems in the quark sector where are different observables in the up and down sector. Usually, the mixing and masses pattern are
studied through the flavor changing neutral current processes (FCNC) and constraints on flavor symmetries are obtained \cite{FCNC,FCNC331,analisis331}. In the leptonic sector many analysis
have been done in the neutrino sector using specially the available data coming from different neutrino oscillation experiments \cite{neutrinos}. An important sector of phenomenology
is concern to charged lepton flavor violating processes (LFV) because they provide direct information about the flavor structure of the lepton sector. Phenomenologically,
there are different models and extensions of the SM that explain LFV. One common approach is to include non-renormalizable effective operators of dimension five and six in order
to have a source of  LFV processes \cite{Weinberg:1979sa}. The effective operators approach is quite general to look for the new physics effects at low energies but the limits obtained
on the LFV couplings are not translate easily to specific models \cite{operadores}.  

Experimentallly, there are searches  of charged LFV processes using channels like 
$\mu \to e \gamma$, $\tau  \to e(\mu) \gamma$, $\mu \to eee$ and conversion $e-\mu$ nuclei \cite{Deppisch:2012vj}.
The experiment MEG, PSI (Switzerland) gives the best limits on $\mu^+\to e^+\gamma$, reporting ${\rm Br}(\mu^+\to e^+\gamma)< 2.4\times 10^{-12}$
and they expect to reach a sensibility $\approx {\rm few}\times 10^{-13}$. Similarly, for the process $\mu^+\to e^+e^+e^-$ is reported by SINDRUM I to be  in the range of $10^{-12}$
 and  the next collaborations MuSIC and $\mu3e$ expect to reach up to $\approx  10^{-16}-10^{-15}$. And for the processes $\mu-e$ nuclei conversion, the rate reported
 by  SINDRUM II  is ${\rm R}_{\mu\,e}(Au)< 7\times 10^{-13}$ and the future experiments Mu2e (FNLA) and COMET (J-PARC) expecto to reach ${\rm R}_{\mu\,e}(Al)\approx  10^{-16}$.
On the other hand, LFV processes have been also searched in the tau lepton sector and around 48 decay channels have been studied by BABAR and BELLE \cite{Asner:2010qj}. The best results
obtained at $90 \%$ C.L. are ${\rm Br}(\tau\to e\gamma)<3.3\times 10^{-8}$,  ${\rm Br}(\tau\to \mu\gamma)<4.4\times 10^{-8}$ and three body decays
${\rm Br}(\tau^-\ell_1^-\ell_2^+\ell_3^-)\lesssim (1.5-3.0)\times 10^{-8}$ with $l=e, \mu$ \cite{Adam:2011ch}. Future experiments like SuperB or BELLE II could get sensibilities of the order
of $10^{-9}$. And finally, one new player in the experimental enviroment is arriving and it is LHCb, they have reported an upper bound 
on ${\rm Br}(\tau^-\to\mu^-\mu^-\mu^+)<6.3\times 10^{-8}$ \cite{Giffels:2008ar}.

One of the motivations of physics beyond the SM is to solve the flavor problem and explain the patterns and mixings in the fermion phenomenology. In grand unified theories,
different representations are assigned to the fermions and therefore different patterns can emerge in the quark sector as well as the lepton sector. LFV processes in  the framework 
of the extended gauge theories is one option to test these models.
One possible
alternative is based on the gauge symmetry  $SU(3)_c\otimes SU(3)_L\otimes U(1)_X$, known as 331 models\cite{331}. These
models can explain why there are three fermionic families through
the chiral anomaly cancellation condition and the number of colors in QCD. On the other hand, the models based on the 331 symmetry are built in such 
a way that the couplings of the quarks
 with the new neutral $Z'$ boson are not universal in the interaction basis, making them
  not diagonal  in the mass eigenstates basis and yielding to flavor changing neutral currents (FCNC) at tree level \cite{FCNC331}. This is a special feature of the 331 models,
  due to
one quark family being in a different representation of the gauge group to the other two families, in order to satisfy the chiral anomaly cancellation condition. It 
is worth  mentioning that in some
331 models there are not only contributions from the left handed neutral current but also from the right handed neutral currents.
 There are many studies of these new FCNC in the quark sector where are different observables in the up and down sectors 
 that constrain such kind of processes\cite{FCNC331}. In contrast, there are not so many analysis in the leptonic sector, where
leptonic flavor violation (LFV) processes at tree level are present. In particular, LFV processes such as $\tau\rightarrow l^-l^+l^-$ with $l=e,\mu$, have been discussed in the framework
of the minimal supersimmetric standard model, Little Higgs models, left-right symmetry models and many other extensions
of the SM \cite{lfvmodels}. Some of these models predict branching fractions 
for $\tau\rightarrow l^-l^+l^-$ of the order of $10^{-7}$ which could be  detected in
 future experiments.

Different 331 models can be built \cite{Diaz:2003dk}, 
they can be distinguished using the electric charge of the new particles introduced in the spectrum and the structure of
the scalar sector, where models without exotic charges will be considered.
In general, the 331 models are classified depending on how they cancel the chiral anomalies: there are two models that cancel out the anomalies requiring 
just one family and eight models where the three families are required. In the three family models, there are four models where the
leptons are treated identically, two of them treat two quark generations identically and finally, there are two models where all the lepton generations
are treated differently  \cite{Diaz:2003dk}. There is one of these 331 models where the leptonic
sector is described by five left handed leptonic triplets in different representations of the $SU(3)_L$ gauge group. Using these five leptonic representations it
is possible to obtain models where the
 three known leptons coupled to the $Z'$ boson are very different with respect to the new ones. We concentrate on these  models  
 in this work,   studying the
LFV processes and obtain
constraints on the leptonic mixing matrix. In the next section we are going to present 
the main features of the model under consideration and then we focus on the LFV processes, namely
$\tau\rightarrow l^-l^+l^-$ with $l=e,\mu$, and $\mu \to 3 e$, $\mu \to e \gamma$ and $\tau \to \mu (e) \gamma$.

\section{The Model 331}

The model considered  is based on the local gauge symmetry $SU(3)_C\otimes SU(3)_L \otimes U(1)_X$ (331),
where it is common to write the electric charge generator as a linear
combination of the diagonal generators of the group:
 \begin{equation}
Q\ = \ T_3\,+\,\beta\, T_8\,+\,X\ ,
\end{equation}
where the parameter $\beta$ is used to label the particular type of 331 model considered. For constructing the model
we choose $\beta= -1/\sqrt{3}$, which corresponds to models where the new fields in the spectra do not have exotic electric charges.\\

The quark content of this model is described by
\begin{eqnarray}
q_{mL} = \begin{pmatrix}
u_m \cr d_m \cr B_m\end{pmatrix}_L\, \sim
(3,3,0),&\qquad& q_{3L}=\begin{pmatrix}
d_3 \cr u_3 \cr T_3
\end{pmatrix}_L\,\sim(3,3^*,1/3)\nonumber\\
d^c\sim (3^*,1,1/3),\qquad u^c\sim (3^*,1,-2/3),&\qquad& B_m^c\sim (3^*,1,1/3),\qquad T^c\sim (3^*,1,-2/3) ,
\end{eqnarray}
where $m=1,2$ and their assigned quantum numbers of $SU(3)_C\otimes SU(3)_L \otimes U(1)_X$ are shown in the parenthesis.

For the leptonic spectrum we use
\begin{eqnarray}
\Psi_{nL}=\begin{pmatrix}e_n^-\cr \nu_n \cr N_n^0
\end{pmatrix}_L\,\sim (1,3^*,-1/3)
,&\qquad& \Psi_{L}=\begin{pmatrix}
\nu_1 \cr e_1^-\cr E_1^-
\end{pmatrix}_L\,\sim (1,3,-2/3),\nonumber\\
\Psi_{4L}=\begin{pmatrix}
E_2^-\cr N_3^0\cr N_4^0
\end{pmatrix}_L\sim (1,3^*,-1/3), &\qquad& \Psi_{5L}=\begin{pmatrix}
N_5^0\cr E_2^+\cr e_3^+
\end{pmatrix}_L\sim (1,3^*,2/3),\nonumber\\
e_n^c\sim (1,1,1),\qquad e_1^c\sim (1,1,1),&\qquad& E_1^c\sim (1,1,1),\qquad E_2^c\sim (1,1,1),
\end{eqnarray}
\noindent with $n=2,3$. The
five leptonic triplets together  with the quark content insures cancellation of chiral anomalies\cite{analisis331}.
Furthermore, notice that with this proposed assemble for the leptonic sector,
there is only one of the triplets that is not written in the adjoint representation of $SU(3)_L$
and it contains one of the standard lepton families of the SM.

On the other hand, in 331 models without exotic charges, the gauge bosons of the $SU(3)_L$ which
transform according to the adjoint representation, are given by
\begin{equation}
\mathbf{W}_\mu \ = \ W_\mu ^a \frac{\lambda^a}{2} \ = \ \frac 12\left(
\begin{array}{ccc}
W_\mu^3 + \frac 1{\sqrt{3}}W_\mu ^8 & \sqrt{2}\, W_\mu ^{+} &
\sqrt{2}\, K_{1\mu}^0 \\
\sqrt{2}\, W_\mu ^{-} & -W_\mu ^3 + \frac 1{\sqrt{3}} W_\mu ^8 &
\sqrt{2}\, \bar K_{2\mu}^+ \\
\sqrt{2}\, \bar K_{1\mu}^0 & \sqrt{2}\, K_{2\mu}^- & -\frac 2
{\sqrt{3}}W_\mu ^8
\end{array}
\right) \ ,
\label{3}
\end{equation}
where $\lambda^a$  are the Gell-Mann matrices for the considered group. The gauge boson field
$B_\mu$ is associated with the  $U(1)_X$ group which is a singlet under  $SU(3)_L$ and it does not have electric charge.
Once the gauge boson sector is identified then the bosons of the neutral sector  $(W^3,
W^8, B)$ are rotated to get the new neutral gauge bosons $A$, $Z$ and $Z'$: 
\begin{eqnarray}
\left(\begin{array}{c}
A \\ Z \\ Z^\prime
\end{array}\right) =
\left( \begin{array}{ccc}
S_W & - S_W/\sqrt{3} & C_W\sqrt{1- T_W^2/3} \\
C_W &  S_W T_W/\sqrt{3} & -\,S_W\sqrt{1- T_W^2/3} \\
0 & -\sqrt{1- T_W^2/3} &  -T_W/\sqrt{3}
\end{array} \right)
\left( \begin{array}{c}
W^3 \\ W^8 \\ B
\end{array} \right) \ ,
\label{bosgauneu}
\end{eqnarray}
where $\theta_W$ is the Weinberg's angle defined by $T_W = \tan\theta_W
= g'/\sqrt{g^2+{g'}^2/3}$, with $g$ and $g'$ the coupling constants of the
$SU(3)_L$ and $U(1)_X$ groups respectively ($S_W = \sin\theta_W$, $C_W = \cos\theta_W$).
In this new basis, the photon  $A_\mu$ is the gauge boson associated to the charge  generator $Q$ while the $Z_\mu$ boson can be identified as the usual $Z$ gauge boson 
of the SM.  
  Scalar states in these models in general can be considered as real fields, therefore
 the neutral heavy state
$\sqrt{2}\,$Im$K$ decouples from the other neutral bosons, becoming an exact mass eigenstate.
However, the vector bosons $Z$, $Z'$ and $\sqrt{2}\,$Re$K$ in general mix
\cite{Diaz:2003dk}. Then, one can rotate to
the mass eigenstate basis, say $Z_1$, $Z_2$, $Z_3$ (where $Z_1$ is the
ordinary gauge boson seen in high energy experiments) through an orthogonal
mixing matrix $R$:
\begin{equation}
\left(\begin{array}{c}
Z \\ Z' \\ \sqrt2\,{\rm Re}K \end{array}
\right) \ = \ R\;
\left(\begin{array}{c}
Z_1 \\ Z_2 \\ Z_3 \end{array}
\right) \ .
\end{equation}
For the purpose of this work we will assume that ${\rm Re}K$ does not mix with the $Z$ and $Z'$ bosons.\\
About the scalar sector, we are going to break the symmetry in such a way that
\begin{equation}
 SU(3)_c\otimes SU(3)_L \otimes U(1)_X \rightarrow SU(3)_c\otimes SU(2)_L \otimes U(1)_X 
 \rightarrow SU(3)_c \otimes U(1)_Q,
\end{equation}
and we use the following three scalars $\phi_1(1,3^*,-1/3)$ with $<\phi_1>=(0,0,V)^T$, $\phi_2(1,3^*,-1/3)$ with $<\phi_2>=(0,v\sqrt{2},0)^T$,
$\phi_3(1,3^*,2/3)$ with $<\phi_3>=(v'/\sqrt{2},0,0)^T$, and $V>v\sim v'$ \cite{analisis331}.

Our main aim  concerns  the leptonic phenomenology and therefore only the leptonic sector will be addressed.
The Lagrangian for the neutral currents in this sector is
\begin{eqnarray}\label{lagrangiano}
{\cal L}_{NC}&=&-\sum_\ell\left[g S_W\, A_\mu\left\{\bar{\ell^0}\gamma_\mu\epsilon^{A}_{\ell_{L}}P_L{\ell}^0
+\bar{\ell^0}\gamma_\mu\epsilon^{A}_{\ell_{R}}P_R{\ell}^0 \right\}\right.\nonumber\\
&&\hskip1.cm+\frac{gZ^\mu}{2C_W}\left\{\bar{\ell^0}\gamma_\mu\epsilon^{Z}_{\ell_{L}}P_L{\ell}^0
+\bar{\ell^0}\gamma_\mu\epsilon^{Z}_{\ell_{R}}P_R{\ell}^0 \right\}\nonumber\\
&&\hskip1.cm+\left.\frac{g^\prime Z^{\prime\mu}}{2\sqrt{3}S_W C_W}\left\{\bar{\psi^0}\gamma_\mu\epsilon^{Z'}_{\ell_{L}}P_L{\ell}^0
+\bar{\ell^0}\gamma_\mu\epsilon^{Z'}_{\ell_{R}}P_R{\ell}^0 \right\}\right] ,
\end{eqnarray}
where $\ell^0$ in this notation stands for the charged leptons vector ${ \ell^{0\,T}}=\left(e_1^{0-},  e_2^{0-},  e_3^{0-}, E_1^{0-},  E_2^{0-}\right)$. The 
zero superscript
 denotes that the fields are in the interaction basis, and the couplings to the neutral bosons are 
\begin{eqnarray}\label{acople}
\epsilon_{ { \ell}_{L}}^{A} &=& I_{5\times5} \, ,\nonumber\\
\epsilon_{{\ell}_{(R)}}^{A}&=& I_{5\times5} \nonumber \\
\epsilon_{{ \ell}_{L}}^{\it Z} &=& Diag(C_{2W},C_{2W},C_{2W},-2S_W^2,C_{2W})\, ,\nonumber\\
\epsilon_{{ \ell}_{R}}^{\it Z} &=& Diag(-2S_W^2,-2S_W^2,-2S_W^2,-2S_W^2,C_{2W})\, ,\nonumber\\
\epsilon_{{ \ell}_{L}}^{\it Z'} &=&  Diag(1,-C_{2W},-C_{2W},-C_{2W},-C_{2W})\, , \nonumber\\
\epsilon_{{ \ell}_{R}}^{\it Z'} &=&  Diag(2S_W^2,2S_W^2,-C_{2W},2S_W^2,1)\, ,
\end{eqnarray}
 where $C_{2W}=\cos{(2\theta_W)}$. Notice that the couplings of the standard charged leptons to the photon
$A_\mu$ are universal as well as the couplings to the $Z$ boson. A feature of this model is that the couplings
of the standard left handed leptons as well as the right handed leptons to the $Z'$ boson are not universal,
 due to the fact that one of the lepton triplets is in a different representation to the other two. Since these couplings to the $Z'$ boson are not universal,
at least for the standard leptons, when they are rotated to mass eigenstates the obtained mixing matrix  will allow LFV at tree level.

 A similar procedure in the neutral leptonic sector can be done, 
 ${ N^{0\,T}}=\left(\nu_1^0,  \nu_2^0,  \nu_3^0,  N_1^0,  N_2^0,  N_3^0,  N_4^0, N_5^0\right)$ generating the couplings
\begin{eqnarray}
\epsilon_{{ N}_{L}}^{A}&=&0  \, \nonumber\\
\epsilon_{{ N}_{L}}^{Z} &=& Diag(1,1,1,0,0,1,0,-1)\,  \nonumber \\
\epsilon_{{ N}_{L}}^{Z'} &=& Diag(1,-C_{2W},-C_{2W},2C_W^2,2C_W^2,-C_{2W},2C_W^2,-1) .
\end{eqnarray}
Here the couplings of the standard neutrinos to the photon $A$ and $Z$ boson are universal but the couplings of the corresponding leptons to the $Z'$ are not.\\

It is possible to re-write the neutral current Lagrangian in order to
use the formalism presented in reference \cite{Langacker:2000ju} and generate an effective Lagrangian like
\begin{equation}
{\cal L}_{\rm NC}^{eff} \ = \ - \; e\, J_{em}^\mu\,A_\mu\; -
\; g_1\,J^{(1)\mu}\,Z_{1\mu}\; - \; g_2\,J^{(2)\mu}\,Z_{2\mu}  \ ,
\end{equation}
where the currents associated to the gauge  $Z$ and $Z'$ bosons are
\begin{eqnarray}
J^{(1)}_\mu & = & \sum_{ij} \bar \ell_i^0\, \gamma_\mu\, (\epsilon^Z_{\ell_{L}}
\, P_L + \epsilon^Z_{\ell_{R}} \, P_R) \ell_j^0 \ \ , \\
J^{(2)}_\mu & = & \sum_{ij} \bar \ell_i^0\, \gamma_\mu\, (\epsilon^{Z'}_{\ell_{L}}
\, P_L + \epsilon^{Z'}_{\ell_{R}} \, P_R) \ell_j^0\ \ ,
\end{eqnarray}
with $g_1 = g/C_W$. The $\ell_i^0$ leptons and the gauge bosons  $Z_1$ and $Z_2$ are interaction eigenstates and
 the matrices  $\epsilon^{Z}_{\ell_{L,R}}$ and $\epsilon^{Z'}_{\ell_{L,R}}$ in the charged sector were defined
in equation ~\eqref{acople}. When the fields of the theory are rotated to mass or physical eigenstates the effective Lagrangian for the charged leptons 
can be finally written as:
\begin{equation}
{\cal L}_{\rm eff} \ = \ -\;\frac{4\,G_F}{\sqrt{2}} \sum_{ijkl} \sum_{XY}
\; C^{ijkl}_{XY}\ (\overline{\ell}_i\,\gamma^\mu\, P_X\,\ell_j)\
(\overline{\ell}_k\,\gamma_\mu\, P_Y\,\ell_l)\ ,
\end{equation}
where $X$ and $Y$ run over the chiralities $L,R$ and indices $i,j,k,l$ over the leptonic families. The coefficients $C^{ijkl}_{XY}$ for the standard 
leptons, assuming a mixing angle  $\theta$
between $Z$ and $Z'$ bosons,  are given by \cite{Langacker:2000ju},
\begin{equation}
C^{ijkl}_{XY} \ =
z\,\rho\,\bigg(\frac{g_2}{g_1}\bigg)^2 \,B^{X}_{ij}\, B^{Y}_{kl}\ ,
\label{cijkl}
\end{equation}
where
\begin{eqnarray}
\rho \ & = & \ \frac{m_W^2}{m_{Z'}^2 C^2_W}\,, \nonumber \\
z & = & \left(\sin^2\theta + \frac{m_Z^2}{m_{Z'}^2} \cos^2\theta\right)\nonumber \,,\\
\left(\frac{g_2}{g_1}\right)^2 &= & \frac{1}{3(1-4S_W^2)} \, .
\label{zz}
\end{eqnarray}
The  $B^X$ corresponds to the
 matrices obtained when the unitary matrices  $V^\ell_{L,R}$ are introduced to obtain the mass eigenstates and to diagonalize the Yukawa coupling matrices:
\begin{equation}
{B^X}_{} \ = \ {V_X^{\ell\,\dagger}\;\epsilon_\ell^{Z'}\; V_X^\ell}_{}\ .
\label{bes}
\end{equation}

For the matrix $V$ we will use a well accepted Ans\"{a}tz \cite{Branco:2004ya} where
\begin{equation}
V_L^\ell \ = \ P \; \tilde V\; K
\label{vdef}
\end{equation}
with $P = {\rm diag}(e^{i\phi_1},1,e^{i\phi_3})$, $K={\rm
diag}(e^{i\alpha_1},e^{i\alpha_2},e^{i\alpha_3})$, and the unitary matrix  $\tilde V$ can be parameterized
using three standard mixing angles $\theta_{12}$, $\theta_{23}$ and $\theta_{13}$ and a phase $\varphi$,
\begin{equation}
\tilde V \ = \left(
\begin{array}{ccc}
c_{12}\,c_{13} & s_{12}\,c_{13} & s_{13}\,e^{-i\varphi} \\
-s_{12}\,c_{23}-c_{12}\,s_{23}\,s_{13}\,e^{i\varphi} &
c_{12}\,c_{23}-s_{12}\,s_{23}\,s_{13}\,e^{i\varphi} &
s_{23}\,c_{13} \\
s_{12}\,s_{23}-c_{12}\,c_{23}\,s_{13}\,e^{i\varphi} &
-c_{12}\,s_{23}-s_{12}\,c_{23}\,s_{13}\,e^{i\varphi} &
c_{23}\,c_{13}
\end{array}
\right)\ \ .
\label{vparam}
\end{equation}
Notice that if we are considering only the standard charged leptons, the coupling matrices in ec.\ref{acople} can be written as
\begin{eqnarray}
\label{eps2}
\epsilon_{{ \ell}_{L}}^{\it Z'} &=&  -(1-2S_W^2) {\bf I}_{3\times 3}+2\,C_W^2\, Diag(1,0,0)\, , \nonumber\\
\epsilon_{{ \ell}_{R}}^{\it Z'} &=&  2\,S_W^2\,{\bf I}_{3\times 3}-Diag(0,0,1)\, .
\end{eqnarray}
The terms which are proportional to the identity are not contributing
to the LFV processes at tree level, while the second term in the above equations does. These equations (\ref{eps2}) correspond to the case where the first family is in
the adjoint representation. However, if the second family was the chosen one to be in a different representation then
the only change is in the second term which is proportional to $Diag(0,1,0)$. Finally,  if instead of that the third family was  chosen,
then again the only change is the position of the entry different from zero in the second term. We should emphasize that the source of LFV
in neutral currents mediated by the $Z'$ boson, comes from the non-diagonal elements
in the $3\times 3$ matrices $B^\ell_{L,R}$.

\section{LFV processes}

Our next task is to get bounds on the parameters involved in the LFV couplings and it is done considering different LFV processes.
Recently, the BELLE \cite{datosexperimentalesbelle}
and BABAR\cite{datosexperimentalesbabar} collaborations have reported measurements of various
LFV channels and they have put new bounds on these branching fractions, see table \ref{tab1}. Other channels to consider
are $BR(\mu^-\rightarrow e^-\gamma)< 2,4 \times 10^{-12}$ \cite{datosexperimentalesbelle} and
$BR(\mu^-\rightarrow e^-e^-e^+)< 1,0 \times 10^{-12}$ \cite{datosexperimentalesbelle}.

\begin{table}
 \centering
\begin{tabular}{|c|c|c|c|}
\hline\hline
Processes & $BR(\times 10^{-8})$ BELLE & $BR(\times 10^{-8})$ BABAR  \\\hline
$\tau^-\rightarrow e^-\gamma$ & 12 & 3.3 \\\hline
$\tau^-\rightarrow \mu^-\gamma$ & 4.5 & 4.4 \\\hline
$\tau^-\rightarrow e^-e^+e^-$ & 2,7 & 2,9 \\\hline
$\tau^-\rightarrow \mu^-\mu^+\mu^-$ & 2,1 & 3,3 \\\hline
$\tau^-\rightarrow e^-\mu^+\mu^-$ & 2,7 & 3,2\\\hline
$\tau^-\rightarrow\mu^-e^+e^-$ & 1,8 & 2,2 \\\hline
$\tau^-\rightarrow e^+\mu^-\mu^-$ & 1,7 & 2,6\\\hline
$\tau^-\rightarrow \mu^+e^-e^-$ & 1,5 & 1,8 \\\hline
\end{tabular}
\caption{Experimental data and their bounds from BELLE  \cite{datosexperimentalesbelle} and BABAR \cite{datosexperimentalesbabar}}
\label{tab1}
\end{table}

\begin{figure}[htp]
 \begin{center}
\includegraphics[width=5cm]{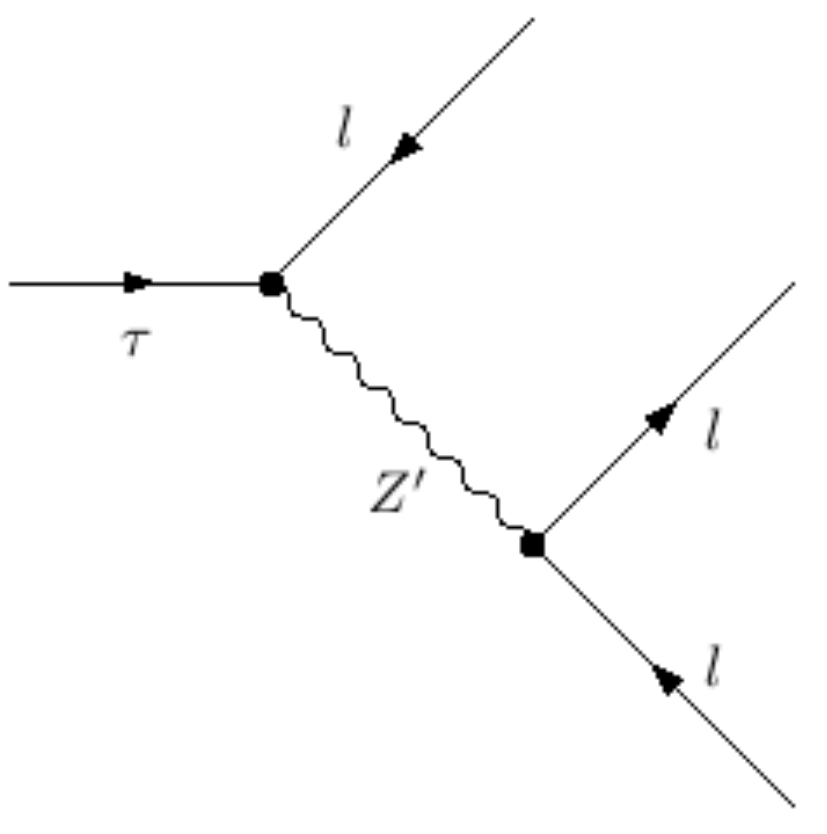}
\includegraphics[width=5cm]{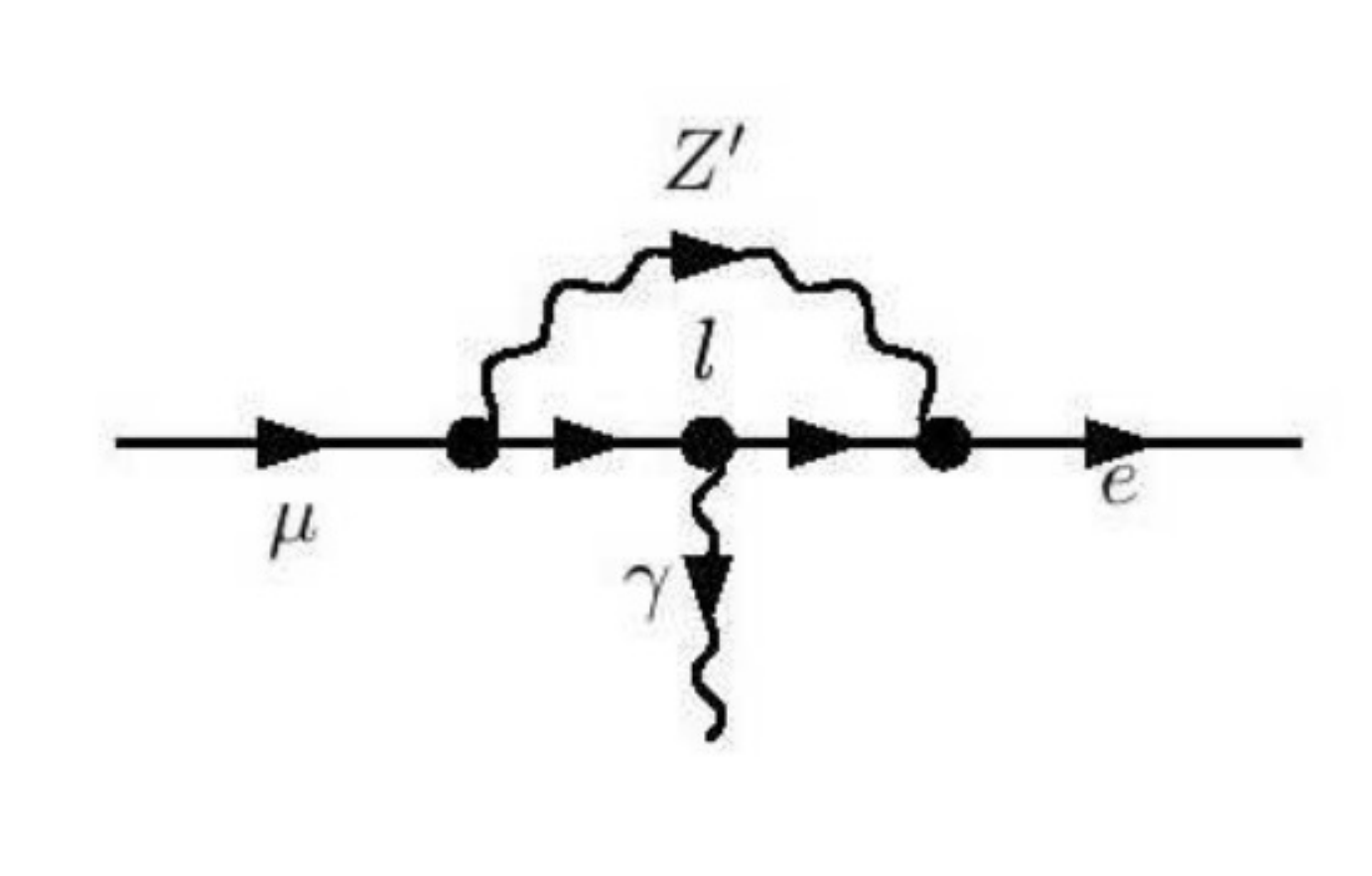}
\end{center}
\caption{Feynman diagram  for the tau(muon) lepton decay into three body through the new $Z'$ current and the diagram for $\mu^- \to e^- \gamma$.}
\label{diagram}
\end{figure}


In the framework of the model 331 that we  presented in section 2, we calculated the decay widths for the different processes 
that we are going to consider. For
the $l_j\rightarrow l_i\gamma$ processes, the decay widths are

\begin{eqnarray}
\Gamma(l_j\rightarrow l_i\gamma) = && \frac{\alpha G_F^2 M_j^3}{8\pi^4} \left(\frac{g_2}{g_1}\right)^4\rho^2
 \left[\left(B^R M_l B^L\right)^2_{ij}+ \left(B^L M_l B^R\right)^2_{ij}  \right]\nonumber\\ ,
 \label{decays1}
\end{eqnarray}
with $i,j=e,\mu,\tau$, and $M_l$ a diagonal mass matrix where the electron mass has been neglected.
 From table 1, we should also evaluate the decay widths into three charged leptons, see figure \ref{diagram},
\begin{eqnarray}
\Gamma(l_j\to l_i^- l_i^- l_i^+ )= &&\frac{G_F^2 M_{l_j}^5}{48 \pi^3}\left(\frac{g_2}{g_1}\right)^4 \rho^2\nonumber\\
&\times&\left[2 \left|B_{ij}^L B_{ii}^L\right|^2+2 \left|B_{ij}^R B_{ii}^R\right|^2+ \left|B_{ij}^L B_{ii}^R\right|^2+
\left|B_{ij}^R B_{ii}^L\right|^2\right] \, ,\nonumber\\
\Gamma(l_j\to l_i^- l_k^- l_l^+ )= &&\frac{G_F^2 M_{l_j}^5}{48 \pi^3}\left(\frac{g_2}{g_1}\right)^4 \rho^2\nonumber\\
&\times&\left[ \left|B_{ij}^L B_{kl}^L+B_{kj}^L B_{il}^L\right|^2+ \left|B_{ij}^R B_{kl}^R+B_{kj}^R B_{il}^R\right|^2+
\left|B_{ij}^L B_{kl}^R\right|^2+\left|B_{kj}^L B_{il}^R\right|^2 \right.\nonumber\\
&& +\left.
\left|B_{ij}^R B_{kl}^L\right|^2+\left|B_{kj}^R B_{il}^L\right|^2\right]\nonumber\\ ,
\label{decays2}
\end{eqnarray}
where the elements $B_{ij}^{L,R}$ are defined in equation (\ref{bes}) and $\rho$ in equation (\ref{zz}).

In order to do the numerical analysis, we trace back the final parameters which are going
to be present in the decay widths, namely the mixing angles $\theta_{12}$, $\theta_{23}$,  $\theta_{13}$ and the $Z'$ gauge boson mass. There are also
phases coming from the $V_L^{l}$
matrix. We have found their effect to be negligible and therefore we have assumed them equal to zero. We are going
to consider two cases depending on which leptonic family is in a different representation of $SU(3)_L$: the first  or the third leptonic
family. We should mention that
the option of the second leptonic family in a different representation is completely analogous to the case of the first family, so we do not present that case.

\begin{figure}[htp]
 \begin{center}
\includegraphics[width=8cm]{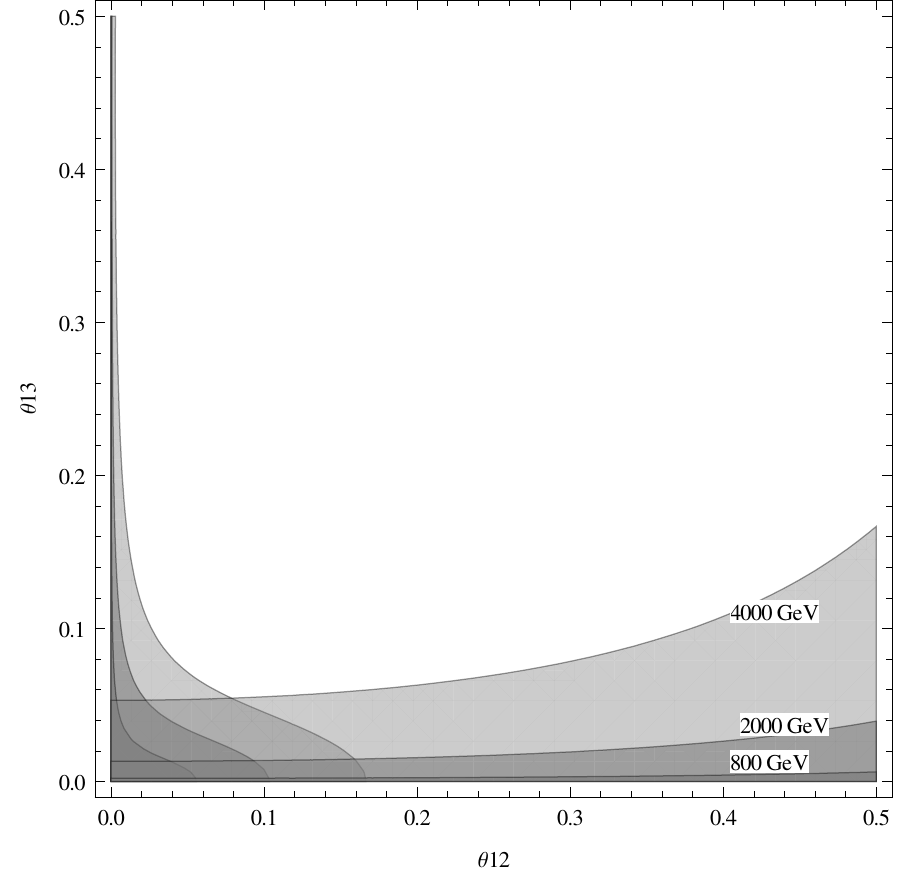}
\end{center}
\caption{The allowed region from processes $\tau\to lll$ and $\mu \to e\gamma$ in the  $\theta_{12}-\theta_{13}$ plane using
different ${Z'}$ boson masses $(800, 2000, 4000)$ GeV.}
\label{P2theta12theta13}
\end{figure}

For the case of the first leptonic family in a different representation, the rotation matrix in the charged leptonic sector depends on
$\theta_{12}$, $\theta_{13}$ and the $Z'$ boson mass, assuming that
the phases involved are zero. Now, we  use the experimental bounds on the different LFV processes shown in table \ref{tab1} in order to obtain constraints 
on the mixing parameters and the $Z'$ boson mass. In general, the observables considered here are proportional to $\rho^2$, equation (\ref{zz}), which is depending on the $Z'$ gauge boson mass resulting in
a dominant factor $\sim m_{Z'}^{-4}$ in equations (\ref{decays1}) and (\ref{decays2}). 
In figure \ref{P2theta12theta13}, bounds coming for the six decay widths of $\tau$ into three charged leptons are shown in the $\theta_{12}-\theta_{13}$ plane, the allowed regions plotted are covering the right
side of the plane. We 
have used $Z'$ boson masses of
  $(800, 2000, 4000)$ GeV. On the other hand, from the process $\tau\to e\gamma$, it is observed that for $\theta_{12}<0.1$ the mixing angle $\theta_{13}$ could be up to
$(0.08, 0.14, 0.2)$ for the $Z'$ boson masses $m_{Z'}=(800, 2000, 4000)$ GeV, it is corresponding to the zones allowed in the left side of the plot. Finally, the processes $\mu\to eee$ and $\tau \to \mu\gamma$ are 
not generating stronger
bounds on the parameters than the ones mentioned previously.

\begin{figure}[htp]
 \begin{center}
\includegraphics[width=5cm]{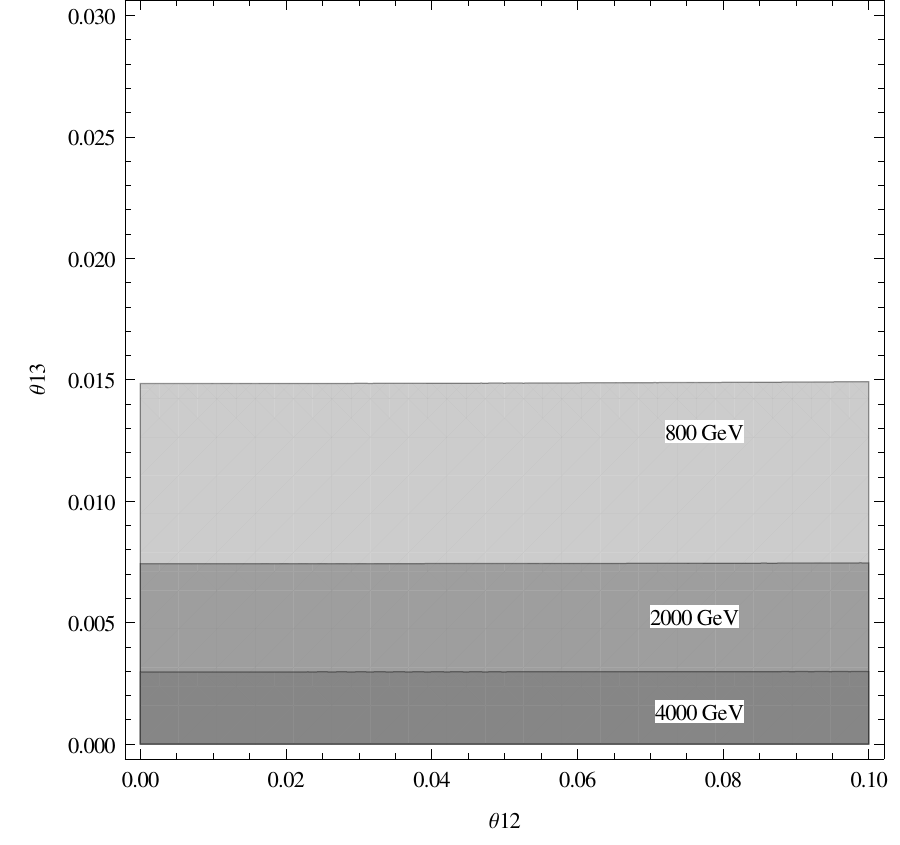}
\includegraphics[width=5cm]{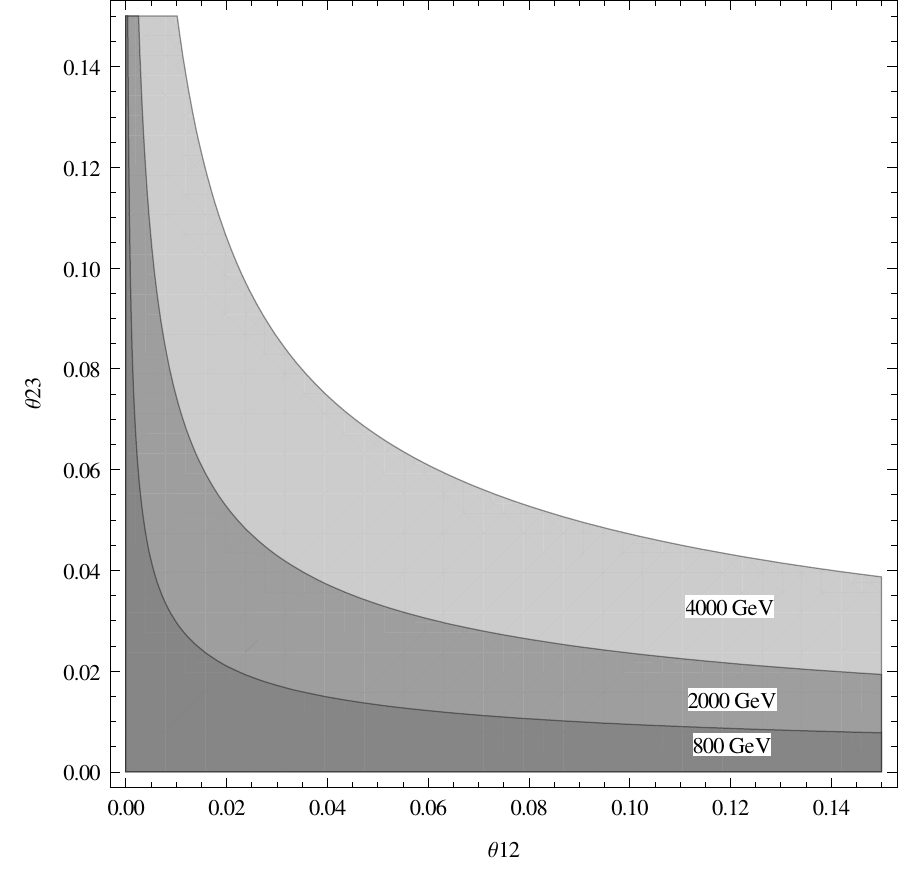}
\includegraphics[width=5cm]{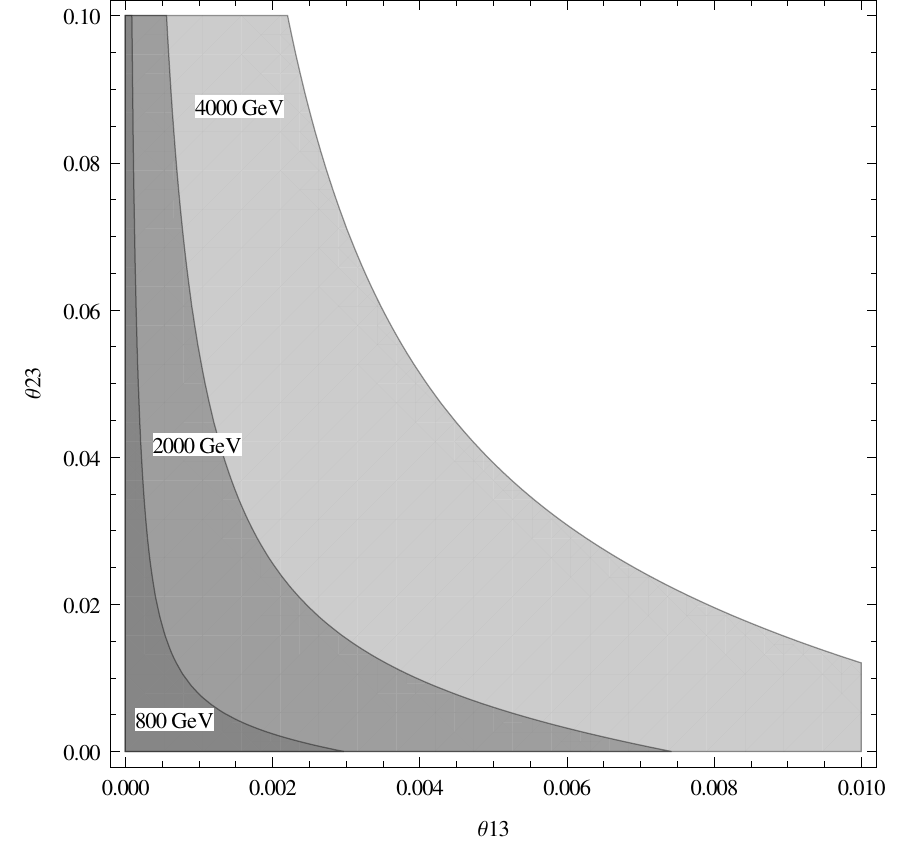}
\end{center}
\caption{Bounds coming from the $\mu\to e\gamma$ process in the different planes such that the third mixing angle is set to zero for $m_{Z'}=(800, 2000, 4000 GeV)$.}
\label{Tmuegamma}
\end{figure}

\begin{figure}[htp]
 \begin{center}
\includegraphics[width=8cm]{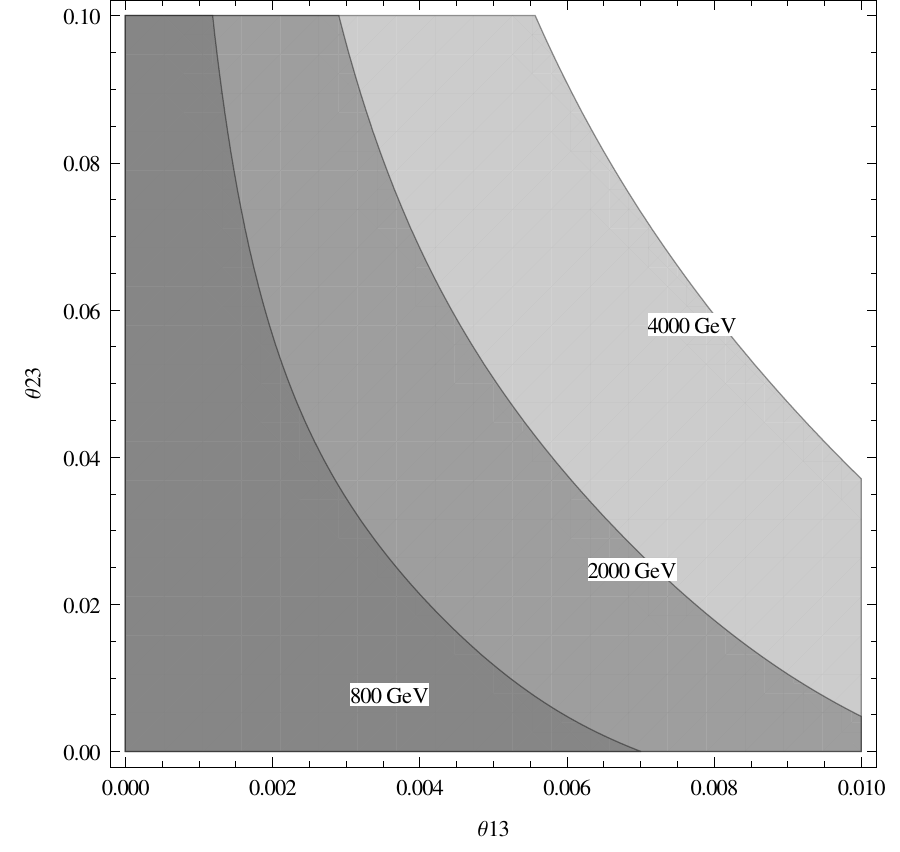}
\end{center}
\caption{Bounds from $\mu\to eee$ in the  $\theta_{13}-\theta_{23}$ plane using
$m_{Z'}=(800, 2000, 4000)$ GeV.}
\label{Tmu3e}
\end{figure}

\begin{figure}[htp]
 \begin{center}
\includegraphics[width=5cm]{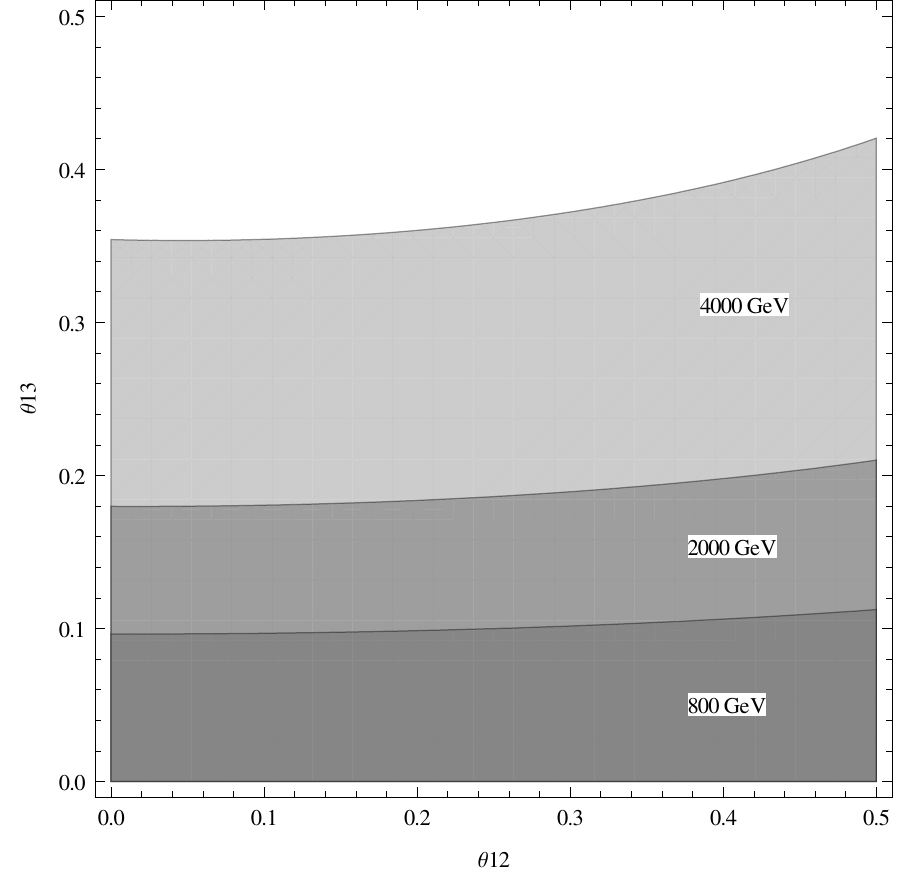}
\includegraphics[width=5cm]{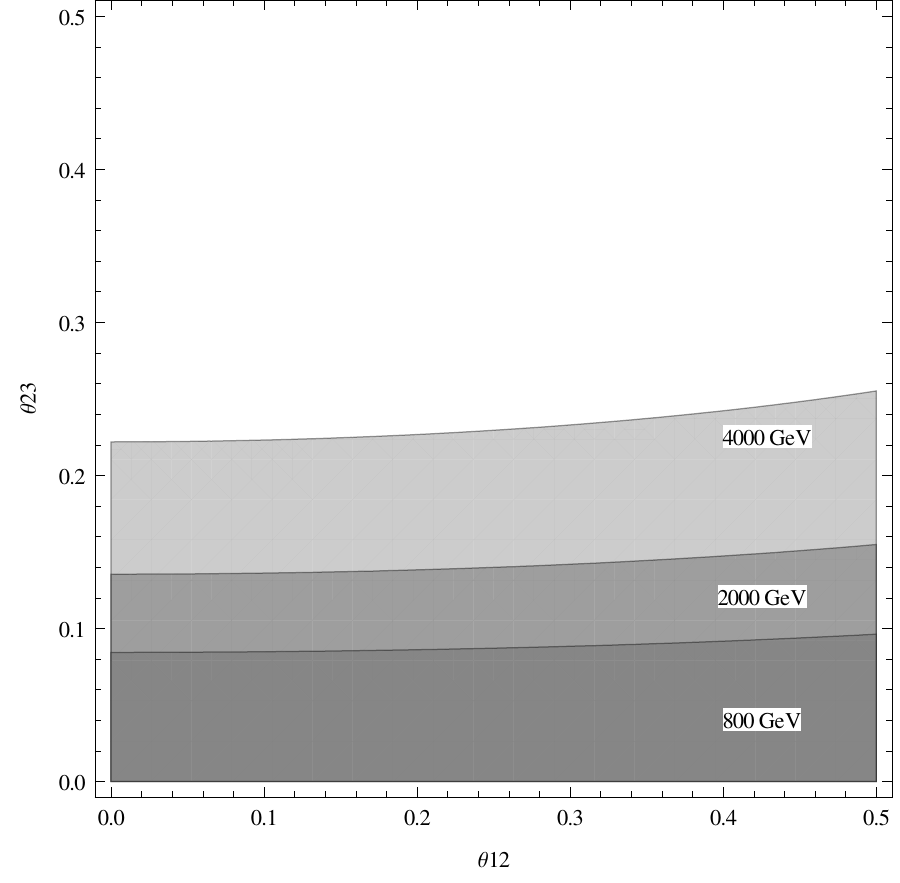}
\includegraphics[width=5cm]{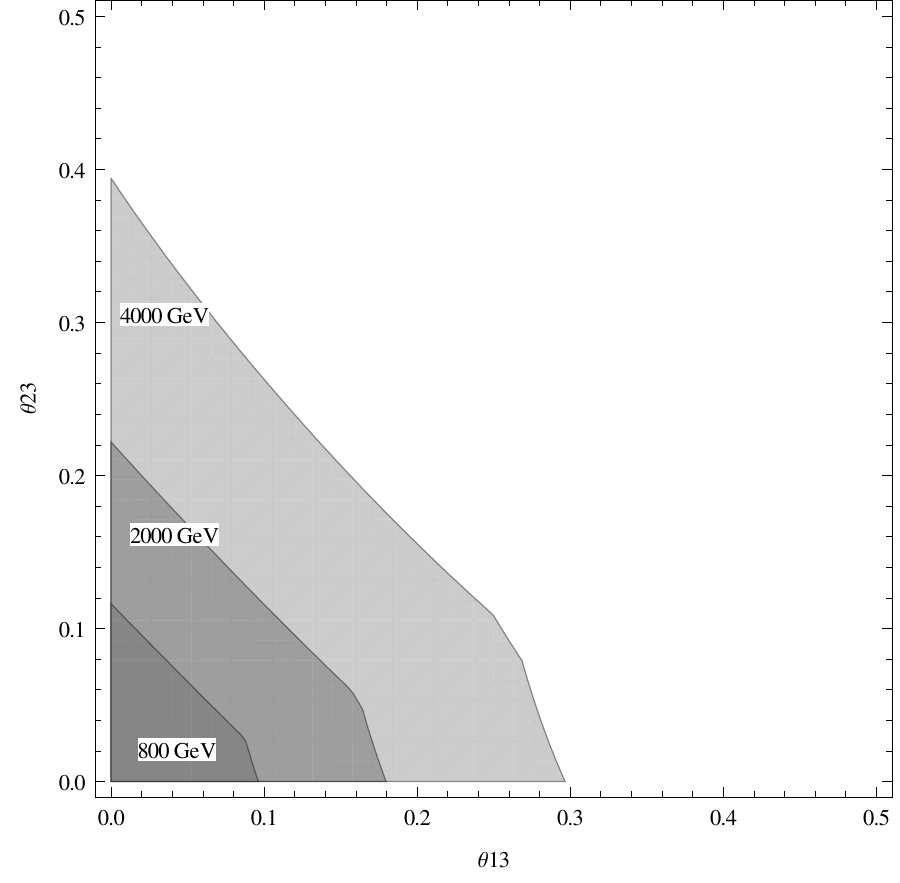}
\end{center}
\caption{Bounds obtained using the $\tau\to lll$ processes of table \ref{tab1}, with the three different scenarios described in section 3
 for  $\theta_{12}$, \, $\theta_{13}$, and $\theta_{23}$ with $m_{Z'}=(800, 2000, 4000)$ GeV.}
\label{Ttau3l}
\end{figure}

\begin{figure}[htp]
 \begin{center}
\includegraphics[width=8cm]{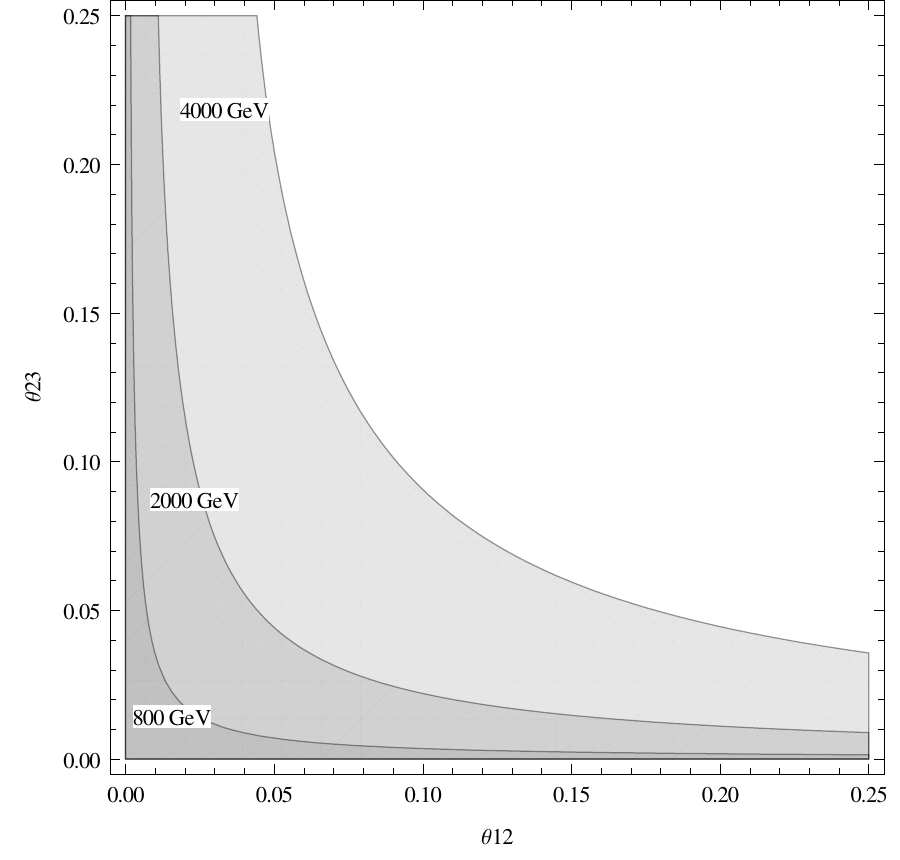}
\end{center}
\caption{Bounds from $\tau\to\mu\gamma$ in the  $\theta_{13}-\theta_{23}$ plane with
$m_{Z'}=(800, 2000, 4000)$ GeV.}
\label{Ttaumugamma}
\end{figure}


Now, the case of the third family transforming differently to the other two families using masses for the $Z'$ gauge boson of $800,2000,4000$ GeV. In Figure \ref{Tmuegamma}, it is shown the results from the process
$\mu \to e \gamma$ with the three mixing angles, we are going to take one of them zero each time and show the plane of the other two angles. In the case of  $\theta_{23}=0$ then the plane  $\theta_{12}-\theta_{13}$ is plotted (left figure), there the mixing angle  $\theta_{12}$ 
does not get any bound in this range from this observable and the angle $\theta_{13}$ should be
of the order of $10^{-2}-10^{-3}$ for the $Z'$ masses considered. Taking $\theta_{13}=0$ then the plane $\theta_{12}-\theta_{23}$ is shown (center figure), the mixing angles  $\theta_{12}-\theta_{23}$ should be of 
the order of $10^{-1}$. And finally considering  $\theta_{12}=0$ then the plane $\theta_{23}-\theta_{13}$ is shown (rigth figure), again
the other two mixing angles are of order of $10^{-1}$. 
In figure \ref{Tmu3e}, the  $\mu \rightarrow eee$  decay is considered. For this decay taking the cases $\theta_{13}=0$ and $\theta_{23}=0$, any improved bound is obtained, assuming small mixing angles. But for 
the case $\theta_{12}=0$, the plane
$\theta_{23}-\theta_{13}$ shown, the mixing
angles are of the order of $\theta_{23}\sim 10^{-2}$ and $\theta_{13}\sim 10^{-3}$. Figure \ref{Ttau3l} is considering the decay $\tau\rightarrow lll$ which are six different decays, see table \ref{tab1}. Following the same analysis, when
$\theta_{23}=0$ there is not a strong depence on $\theta_{12}$ while the mixing angle $\theta_{13}$ is of the order of $\sim 10^{-1}$ (left figure). Similarly with $\theta_{13}=0$, there is not sensitive to $\theta_{12}$  
but $\theta_{23}\sim 10^{-1}$ (center figure).
And when $\theta_{12}=0$ is considered then the other two mixing agles remain in the same order of magnitude and only change with the $Z'$ gauge boson mass (right figure). Finally, in figure \ref{Ttaumugamma} bounds using
the $\tau \rightarrow \mu\gamma$
are obtained. We have noticed that taking $\theta_{12}=0$ and $\theta_{23}=0$ then there is not any improved bounds on the parameters and for the case $\theta_{13}=0$ then the mixing angles $\theta_{12}$ and $\theta_{23}$ are 
in the same order of magnitude. We should mention that using the process $\tau \rightarrow e\gamma$ there is not obtained any bound on the mixing angles lower than the obtained previously, so it is less restrictive than 
the other processes considered here. The whole set of bounds obtained in the figures 2-5 help to obtain the order of magnitude of the mixing angles involved in order to satisfy the experimental bounds on the LFV processes 
considered in this work. There is one order of magnitude of difference between them and the hierarchy $\theta_{12}\sim 10^{-1}>\theta_{23}\sim 10^{-2}>\theta_{13}\sim 10^{-3}$.

Now we can use the information obtained about the mixing angles and see what is happening in the neutrino sector of the model. First of all, we write down the charged current Lagrangian as:
\begin{eqnarray}
{\cal L}_{cc}= -\frac{g}{\sqrt{2}}(\bar{\nu}_L^0\,\gamma^\mu\, \ell_L^0)W_\mu^++{\rm h.c.}
\end{eqnarray} 
where quantum fields are in the weak  interaction basis. In order to go to the mass eigenstates, it is necessary to introduce the rotation matrices  $V_L^{\nu}$ and $V_L^\ell$ which are also diagonalizing the Yukawa matrices. 
These matrices are defining the mixing matrix in the leptonic sector known as PMNS matrix and it is
$V_{PMNS}= V_L^\nu (V_L^\ell)^\dag $. The matrices $V_L^{\nu}$ and $V_L^\ell$ are parameterized as in equation (19). Now taking into account that $\theta_{12}\sim 10^{-1}>\theta_{23}\sim 10^{-2}>\theta_{13}\sim 10^{-3}$, it is possible to obtain
the following limits on the mixing leptonic matrix
\begin{eqnarray}
|V_L^\ell| = \begin{pmatrix} 1\to 0.995004 & 0\to 0.0998334 & 0\to 0.001\cr 0\to 0.0998384  & 1\to 0.994953 & 0\to 0.00999983 \cr 0\to 3.36328\times10^{-6} & 0\to0.0100497 & 1\to 0.99995  
\end{pmatrix} \,\, .
\end{eqnarray}
On the other hand, considering massive neutrinos and  the oscillation data, the mixing angles in the neutrino sector
are around $\theta_{12}\sim 30^o\,, \theta_{23}\sim 45^o$ y $\theta_{13}\sim 8^o$ \cite{Deppisch:2012vj},\cite{GonzalezGarcia:2012sz},\cite{GonzalezGarcia:2007ib},\cite{Tortola:2012te},\cite{Fogli:2012ua} where the final mixing
matrix is (at $3\sigma$ C.L.) 
\begin{eqnarray}
|V_L^\nu|= \begin{pmatrix} 0.795\to 0.846 & 0.513\to 0.585 & 0.126\to 0.178 \cr 
0.205\to 0.543 & 0.416\to 0.730 & 0.579\to 0.808 \cr 0.215\to 0.548 & 0.409\to 0.725 & 0.567\to 0.800
\end{pmatrix} \, \, .
\end{eqnarray}
Finally, we can combine the matrices to obtain the PMNS matrix and it is
\begin{eqnarray}
|V_{PMNS}|= \begin{pmatrix} 0.795\to 0.900 & 0.513\to 0.668 & 0.126\to 0.183 \cr 
0.205\to 0.613 & 0.416\to 0.788 & 0.579\to 0.815 \cr 0.215\to 0.618 & 0.409\to 0.784 & 0.567\to 0.807
\end{pmatrix}
\end{eqnarray}
which is in agreement with the accepted values for this matrix in the literature \cite{Deppisch:2012vj},\cite{GonzalezGarcia:2012sz},\cite{GonzalezGarcia:2007ib},\cite{Tortola:2012te},\cite{Fogli:2012ua}.

\section{Conclusions}

In this work, we have addressed the LFV processes in a model based on the 331 symmetry where the leptonic sector is described by five left handed leptonic triplets in
different representations of the $SU(3)_L$ gauge group. Here, the couplings of the new neutral $Z'$ boson with the usual
leptons are not universal. This feature is due to  one of the
lepton triplets being in a different representation than the other two, which leads to LFV at tree level once they are rotated to  mass eigenstates.  We have considered
 some LFV processes which have been measured by the BELLE and the BABAR collaborations: $\tau \to 3l$,
$\tau \to l \gamma$, $\mu \to e \gamma$ and $\mu \to 3 e$ (see table 1). Tha analysis was done considering two cases depending on 
which leptonic family is in the different
representation
of $SU(3)_L$ in the 331 model described in section 2. For the first case (where the first leptonic family is in the different representation), we obtained allowed regions 
on the  $\theta_{12}-\theta_{23}$ plane of the order of $\sim 10^{-1}$.
For the second case (the third leptonic family in a different representation), the bound on the process $\mu \to e \gamma$ constrain the space of parameters 
to regions around $\theta_{12} \sim 10^{-2}$, $\theta_{23}\sim 10^{-2}$
and $\theta_{13} \sim 10^{-3}$.  We also explored the bounds coming from other LFV processes, which are consistent with these regions 
and the results are shown in figures 2-5. It is worth to point out that the mixing angles obtained for the leptons in the framework of the 331 model considered here is generating a matrix 
which is almost an identity matrix which is according with the no experimental evidence of the FCNC at low energies . Therefore,
even considering the mixing in the neutrino sector the results are in agreement with the experimental values reported for the PMNS matrix.

\section{Acknowledgements}

We are very thankful for  the hospitality of the IFT, Sao Paulo where this work was finished and to Profesor Vicente Pleitez and Prof. C. Sandoval for discussions 
and comments about this work. The works of J.D. and J.A.R. are supported in part by the DIB-UNAL 14844 grant.


\begin{thebibliography}{1}


\bibitem{FCNC}
  J.~L.~Hewett, H.~Weerts, R.~Brock, J.~N.~Butler, B.~C.~K.~Casey, J.~Collar, A.~de Govea and R.~Essig {\it et al.},
  arXiv:1205.2671 [hep-ex].
  Y.~Nir,
  CERN Yellow Report CERN-2010-001, 279-314
  [arXiv:1010.2666 [hep-ph]].
  
  
\bibitem{FCNC331}
  F.~Pisano and V.~Pleitez,  arXiv:hep-ph/930726;5
  J.~Alexis.~Rodriguez and M.~Sher, Phys.\ Rev.\  D {\bf 70} (2004) 117702, [arXiv:0407248];
  A.~E.~Carcamo Hernandez, R.~Martinez and F.~Ochoa,
  Phys.\ Rev.\  D {\bf 73} (2006) 035007 [arXiv:0510421] ; J.~M.~Cabarcas, D.~Gomez Dumm and R.~Martinez,
  J.\ Phys.\ G {\bf 37} (2010) 045001 [arXiv:0910.5700]; M.~A.~Perez, G.~Tavares-Velasco and J.~J.~Toscano,
  Phys.\ Rev.\  D {\bf 69}, 115004 (2004) [arXiv:0402156].
    J.~M.~Cabarcas, J.~Duarte and J-A.~Rodriguez,  
  Adv.\ High Energy Phys.\  {\bf 2012}, 657582 (2012)  [arXiv:1111.0315 [hep-ph]].
  R.~H.~Benavides, Y.~Giraldo and W.~A.~Ponce,
  Phys.\ Rev.\  D {\bf 80} (2009) 113009
  [arXiv:0911.3568 [hep-ph]].
   D. G. Dumm, F. Pisano, and V. Pleitez, Mod. Phys. Lett. \textbf{A9}, 1609 (1994)
  C.~Promberger, S.~Schatt and F.~Schwab,
  Phys.\ Rev.\ D {\bf 75}, 115007 (2007)
  [hep-ph/0702169 [HEP-PH]].

  \bibitem{analisis331}
  W.~A.~Ponce, J.~B.~Florez and L.~A.~Sanchez,
  Int.\ J.\ Mod.\ Phys.\ A {\bf 17}, 643 (2002)
  [hep-ph/0103100]. W.~A.~Ponce, Y.~Giraldo and L.~A.~Sanchez,
  hep-ph/0201133

\bibitem{neutrinos}
  Y.~Fukuda {\it et al.}  [Super-Kamiokande Collaboration],
  Phys.\ Rev.\ Lett.\  {\bf 81} (1998) 1562
  [hep-ex/9807003].
  Q.~R.~Ahmad {\it et al.}  [SNO Collaboration],
  Phys.\ Rev.\ Lett.\  {\bf 89} (2002) 011302
  [nucl-ex/0204009].
  K.~Eguchi {\it et al.}  [KamLAND Collaboration],
  Phys.\ Rev.\ Lett.\  {\bf 90} (2003) 021802
  [hep-ex/0212021].
  D.~V.~Forero, M.~Tortola and J.~W.~F.~Valle,
  Phys.\ Rev.\ D {\bf 86} (2012) 073012
  [arXiv:1205.4018 [hep-ph]].


\bibitem{Weinberg:1979sa}
  S.~Weinberg,
  Phys.\ Rev.\ Lett.\  {\bf 43} (1979) 1566.

\bibitem{operadores}
  M.~Raidal, A.~van der Schaaf, I.~Bigi, M.~L.~Mangano, Y.~K.~Semertzidis, S.~Abel, S.~Albino and S.~Antusch {\it et al.},
  Eur.\ Phys.\ J.\ C {\bf 57} (2008) 13
  [arXiv:0801.1826 [hep-ph]].
  L.~Calibbi, Z.~Lalak, S.~Pokorski and R.~Ziegler,
  JHEP {\bf 1207} (2012) 004
  [arXiv:1204.1275 [hep-ph]].

\bibitem{Deppisch:2012vj}
  F.~F.~Deppisch,
  Fortsch.\ Phys.\  {\bf 61} (2013) 622
  [arXiv:1206.5212 [hep-ph]].
  A.~Abada,
  Comptes Rendus Physique {\bf 13} (2012) 180
  [arXiv:1110.6507 [hep-ph]].

\bibitem{Asner:2010qj}

  Y.~Amhis {\it et al.}  [Heavy Flavor Averaging Group Collaboration],
  arXiv:1207.1158 [hep-ex].

  

\bibitem{Adam:2011ch}
  J.~Adam {\it et al.}  [MEG Collaboration],
  Phys.\ Rev.\ Lett.\  {\bf 107} (2011) 171801
  [arXiv:1107.5547 [hep-ex]].
  U.~Bellgardt {\it et al.}  [SINDRUM Collaboration],
  Nucl.\ Phys.\ B {\bf 299} (1988) 1.
  J.~L.~Hewett, H.~Weerts, R.~Brock, J.~N.~Butler, B.~C.~K.~Casey, J.~Collar, A.~de Govea and R.~Essig {\it et al.},
  arXiv:1205.2671 [hep-ex].

\bibitem{Giffels:2008ar}
  M.~Giffels, J.~Kallarackal, M.~Kramer, B.~O'Leary and A.~Stahl,
  Phys.\ Rev.\ D {\bf 77} (2008) 073010
  [arXiv:0802.0049 [hep-ph]].


\bibitem{331}
  F.~Pisano and V.~Pleitez,
  Phys.\ Rev.\  D {\bf 46} (1992) 410 [arXiv:9206242];
  P.~H.~Frampton,
  Phys.\ Rev.\ Lett.\  {\bf 69} (1992) 2889;
  J.~C.~Montero, F.~Pisano and V.~Pleitez,
  Phys.\ Rev.\  D {\bf 47} (1993) 2918 [arXiv:9212271];
  R.~Foot, O.~F.~Hernandez, F.~Pisano and V.~Pleitez,
  Phys.\ Rev.\  D {\bf 47} (1993) 4158  [arXiv:9207264].
  R.~Foot, L.~N.~Hoang and T.~A.~Tran, Phys.\ Rev.\  D {\bf 50}, 34 (1994)
  [arXiv:9402243]; J.~T.~Liu and D.~Ng,
  Phys.\ Rev.\  D {\bf 50}, 548 (1994) [arXiv:9401228];
  J.~T.~Liu, Phys.\ Rev.\  D {\bf 50}, 542 (1994) [arXiv:9312312];
  R.~Foot, O.~F.~Hernandez, F.~Pisano and V.~Pleitez, Phys.\ Rev.\  D {\bf 47}, 4158 (1993) [arXiv:9207264].
  M.~B.~Tully and G.~C.~Joshi, Phys.\ Rev.\  D {\bf 64} (2001) 011301
  [arXiv:0011172].
  P.~V.~Dong and H.~N.~Long,
  Int.\ J.\ Mod.\ Phys.\ A {\bf 21}, 6677 (2006)
  [hep-ph/0507155].
   J.~G.~Ferreira, Jr, P.~R.~D.~Pinheiro, C.~A.~d.~S.~Pires and P.~S.~R.~da Silva,
  Phys.\ Rev.\ D {\bf 84}, 095019 (2011)
  [arXiv:1109.0031 [hep-ph]].
    J.~D.~Ruiz-Alvarez, C.~A.~de S.Pires, F.~S.~Queiroz, D.~Restrepo and P.~S.~Rodrigues da Silva,
  Phys.\ Rev.\ D {\bf 86}, 075011 (2012)
  [arXiv:1206.5779 [hep-ph]].
  S.~Profumo and F.~S.~Queiroz,
  arXiv:1307.7802 [hep-ph].
   P.~V.~Dong, H.~T.~Hung and T.~D.~Tham,
  Phys.\ Rev.\ D {\bf 87}, 115003 (2013)
  [arXiv:1305.0369 [hep-ph]].
   C.~Kelso, C.~A.~d.~S.~Pires, S.~Profumo, F.~S.~Queiroz and P.~S.~R.~da Silva,
  arXiv:1308.6630 [hep-ph].
    P.~V.~Dong, T.~P.~Nguyen and D.~V.~Soa,
  Phys.\ Rev.\ D {\bf 88}, 095014 (2013)
  [arXiv:1308.4097 [hep-ph]].
  
  




\bibitem{lfvmodels}
  I.~Cortes Maldonado, A.~Moyotl and G.~Tavares-Velasco,
  Int.\ J.\ Mod.\ Phys.\ A {\bf 26}, 4171 (2011)
  [arXiv:1109.0661 [hep-ph]].
  E.~Nardi,
  arXiv:1112.4418 [hep-ph].
  J.~I.~Aranda, J.~Montano, F.~Ramirez-Zavaleta, J.~J.~Toscano and E.~S.~Tututi,
  arXiv:1202.6288 [hep-ph].
  B.~M.~Dassinger, T.~.Feldmann, T.~.Mannel and S.~Turczyk,
  JHEP {\bf 0710}, 039 (2007)
  [arXiv:0707.0988 [hep-ph]].
  R.~Benbrik, M.~Chabab and G.~Faisel,
  arXiv:1009.3886 [hep-ph].
  D.~Cogollo, A.~V.~de Andrade, F.~S.~Queiroz and P.~Rebello Teles,
  Eur.\ Phys.\ J.\ C {\bf 72}, 2029 (2012)
  [arXiv:1201.1268 [hep-ph]].

  
  

\bibitem{Diaz:2003dk}
T.~A.~Nguyen, N.~A.~Ky and L.~N.~Hoang,
  Int.\ J.\ Mod.\ Phys.\  A {\bf 15}, 283 (2000) [arXiv:9810273];
    M.~B.~Tully and G.~C.~Joshi,
  Int.\ J.\ Mod.\ Phys.\  A {\bf 18}, 1573 (2003) [arXiv:9810282];
    W.~A.~Ponce, Y.~Giraldo and L.~A.~Sanchez,
  Phys.\ Rev.\  D {\bf 67}, 075001 (2003) [arXiv:0210026];
    P.~V.~Dong, L.~N.~Hoang, D.~T.~Nhung and D.~V.~Soa,
  Phys.\ Rev.\  D {\bf 73}, 035004 (2006) [arXiv:0601046].
  M.~Ozer,
  Phys.\ Rev.\  D {\bf 54}, 1143 (1996).


\bibitem{Langacker:2000ju}
  P.~Langacker and M.~Plumacher,
  Phys.\ Rev.\ D {\bf 62}, 013006 (2000)
  [hep-ph/0001204].

  

\bibitem{Branco:2004ya}
  G.~C.~Branco, M.~N.~Rebelo and J.~I.~Silva-Marcos,
  Phys.\ Lett.\  B {\bf 597} (2004) 155
  [arXiv:hep-ph/0403016].

  












\bibitem{datosexperimentalesbelle}
  K.~Hayasaka [Belle Collaboration],
  PoS ICHEP {\bf 2010}, 241 (2010)
  [arXiv:1011.6474 [hep-ex]].
K. Hayasaka et al. (Belle Collaboration), Phys. Lett. B 666, 18 (2008). [arXiv:1010.3746 [hep-ex]].
K. Hayasaka et al. (Belle Collaboration), Phys. Lett. B 687, 139 (2010).[arXiv:1001.3221 [hep-ex]].
 Belle Collaboration. Y. Miyazaki et. al.
 Phys. Lett. B. 660 (2008).
 Babbar Collaboration.
 Phys. Rev. Lett. 92(121801).


\bibitem{datosexperimentalesbabar}
B. Aubert et al. (BABAR Collaboration), Phys. Rev. Lett. 104, 021802 (2010).
B. Aubert et al. (BaBar Collaboration), Phys. Rev. D 81, 111101 (2010).[arXiv:1002.4550 [hep-ex]].
B. Aubert et al. (BaBar Collaboration),
  arXiv:1202.3650 [hep-ex].



  \bibitem{GonzalezGarcia:2012sz}
  M.~C.~Gonzalez-Garcia, M.~Maltoni, J.~Salvado and T.~Schwetz,
  JHEP {\bf 1212} (2012) 123
  [arXiv:1209.3023 [hep-ph]].
\bibitem{GonzalezGarcia:2007ib}
  M.~C.~Gonzalez-Garcia and M.~Maltoni,
  Phys.\ Rept.\  {\bf 460} (2008) 1
  [arXiv:0704.1800 [hep-ph]].

\bibitem{Tortola:2012te}
  D.~V.~Forero, M.~Tortola and J.~W.~F.~Valle,
  Phys.\ Rev.\ D {\bf 86} (2012) 073012
  [arXiv:1205.4018 [hep-ph]].

\bibitem{Fogli:2012ua}
  G.~L.~Fogli, E.~Lisi, A.~Marrone, D.~Montanino, A.~Palazzo and A.~M.~Rotunno,
  Phys.\ Rev.\ D {\bf 86} (2012) 013012
  [arXiv:1205.5254 [hep-ph]].

  
  
  \end{thebibliography}
\end{document}